\documentclass{emulateapj}
\usepackage{natbib}
\usepackage{rotating}
\usepackage{mathtools}
\usepackage{tablefootnote}
\usepackage{color}

\newcommand{\pasa}{Publ. of the Astr. Soc. of Australia}

\begin{document}

\title{A comparative study of two 47~Tuc giant stars with different s-process
enrichment}

\author{M. J. Cordero}
\affil{Astronomisches Rechen-Institut, Zentrum f\"ur Astronomie der Universit\"at Heidelberg, M\"onchhofstrasse 12-14, D-69120 Heidelberg, Germany\\
\and Landessternwarte, ZAH, Heidelberg University, K\"onigstuhl 12, 69117  Heidelberg, Germany }
\email{mjcorde@ari.uni-heidelberg.de}

\author{ C. J. Hansen}
\affil{Landessternwarte, ZAH, Heidelberg University, K\"onigstuhl 12, 69117  Heidelberg, Germany \\
\and  Dark Cosmology Centre, The Niels Bohr Institute, Copenhagen, Denmark}
\email{cjhansen@lsw.uni-heidelberg.de, cjhansen@dark-cosmology.dk}

\author{C. I. Johnson}
\affil{Harvard-Smithsonian Center for Astrophysics, 60 Garden Street, MS-15, Cambridge, MA 02138, USA}
\email{cjohnson@cfa.harvard.edu}

\author{C. A. Pilachowski}
\affil{Astronomy Department, Indiana University Bloomington, Swain West 319, 727 East 3rd Street, Bloomington, IN 47405-7105, USA}
\email{catyp@astro.indiana.edu}

\date{today}

\begin{abstract}

Here we aim to understand the origin of 47~Tuc's La-rich star Lee~4710.  We 
report abundances for O, Na, Mg, Al, Si, Ca, Sc, Ti, V, Cr, Co, Ni, Zn, Y, Zr, 
Ba, La, Ce, Pr, Nd, and Eu, and present a detailed abundance analysis of
two 47~Tuc stars with similar stellar parameters but different slow neutron-capture (s-)process  enrichment. Star Lee~4710 has the highest known La abundance ratio in this 
cluster ([La/Fe] = 1.14), and star Lee~4626 is known to have normal s-process 
abundances (e.g., [Ba/Eu]$<0$).  The nucleosynthetic pattern of elements
with Z$\ga$56 for star Lee~4710 agrees with the predicted yields of a 
$1.3M_{\sun}$ asymptotic giant branch (AGB) star.  Therefore, Lee~4710 may have been enriched by 
mass transfer from a more massive AGB companion, which is compatible with its 
location far away from the center of this relatively metal-rich 
([Fe/H]$\sim-0.7$) globular cluster.  A further analysis comparing the
abundance pattern of Lee~4710 with data available in the literature reveals 
that nine out of the $\sim200$ 47~Tuc stars previously studied show strong 
s-process enhancements that point towards later enrichment 
by more massive AGB stars.

\end{abstract}

\section{Introduction}
\label{sec:intro}

The abundances of neutron-capture elements in globular clusters (GC) provide additional insight, beyond the light elements O, Na, and Al, into the origin of abundance anomalies found among stars in a given cluster. 
In particular, finding GC stars with unusual enhancements of 
elements produced by the s-process\footnote{slow neutron-capture process} 
enables a direct comparison with theoretical yield predictions.  For example,
comparing the ratio of light (first peak) to heavy (second peak) s-process
elements provides information about the mass and metallicity of the stars in
which the s-process elements were produced.  Most s-process elements are 
produced through one of two channels.  The main channel takes place in
$\la8M_{\odot}$ asymptotic giant branch (AGB) stars where neutrons are 
released in a 
$^{12}$C($p,\gamma$)$^{13}$N($\beta^+,\nu)^{13}$C($\alpha,n$)$^{16}$O reaction
occurring in a $^{13}$C-enriched pocket.  The size and efficiency of the
$^{13}$C-pocket, and in turn the outcome of the main s-process, is not 
currently well-constrained, but the $^{13}$C-pocket is likely located in the
He-intershell region and driven by carbon created in the stellar interior
and protons from the outer H-rich layers 
\citep[e.g.,][]{Sneden2008,Kaeppeler2011,Karakas2014}.  Several 
theoretical studies have shown that M$=1-3M_{\odot}$ AGB stars may produce the 
largest fractional yield of main s-process material peaking around 2$M_{\odot}$
\citep[e.g.][]{Bisterzo2010,Bisterzo2014,Cristallo2011, Cristallo2015}. In more massive 
stars, the weak s-process is thought to be the dominant slow 
neutron-capture process, and the neutron source is a
$^{22}$Ne($\alpha,n$)$^{25}$Mg reaction that is only activated at higher 
densities and temperatures 
\citep[T$\ga 2.5 \cdot 10^8$\,K; e.g.,][]{Pignatari2010}.  While the main 
s-process produces a larger fraction of second-peak neutron-capture elements 
(e.g., Ba and La), the weak s-process tends to produce a higher fraction
of first-peak (e.g., Sr, Y, and Zr) s-process elements \citep{Gallino1998,Busso1999,Pignatari2010, Bisterzo2010,Cristallo2011,Karakas2014}.

The GC 47~Tuc is a useful candidate for analyzing neutron-capture production
because it is nearby, massive, metal-rich ([Fe/H]$\sim-0.7$), and has been
spectroscopically studied in great detail.  Previous analyses have found
that 47~Tuc hosts at least three stellar populations that are distinguished
as having unique Na and O abundance ratios \citep[][`C14']{Carretta2013,Cordero2014}.
However, similar to other GCs, 47~Tuc does not exhibit significant abundance
variations of the iron-peak elements 
\citep[][`T14' and references therein]{Thygesen2014}.  Interestingly, although T14 and C14 found the cluster stars to have a mostly homogeneous heavy element composition, identical to the abundances observed in field stars at the same metallicity, T14 and \citet[][`W06']{Wylie2006} found evidence indicating that some RGB and AGB stars may have enhancements of elements produced by the s-process.

In our previous spectroscopic study of 47~Tuc (C14), we found one
s-process rich giant out of a sample of $\sim$160 stars.  In addition to the
s-process rich stars found by W06 and C14, 
\citet{Dorazi2010} have found five Ba-rich stars ([Ba/Fe]$>$0.51) in a sample 
of 1200 GC stars, and \citet{Cohen2005} found an Y-rich star in M13.  Thus,
s-process enhanced stars are relatively rare in monometallic GCs.  
Interestingly, these s-process rich stars tend to be found in the external 
regions of GCs, which could be a consequence of less efficient binary 
disruption, assuming these s-process rich stars formed in binary systems.
Therefore, in the present work we perform a detailed abundance 
analysis of Lee~4710, an s-process enhanced star in 47~Tuc.  We aim to assess
whether or not the star's heavy element composition can be modeled by 
theoretical AGB yields, and also explore possible formation modes.

\section{Data reduction and abundance analysis}
\label{sec:data}

\begin{figure}
  \begin{center}
    \includegraphics[width=0.98\linewidth]{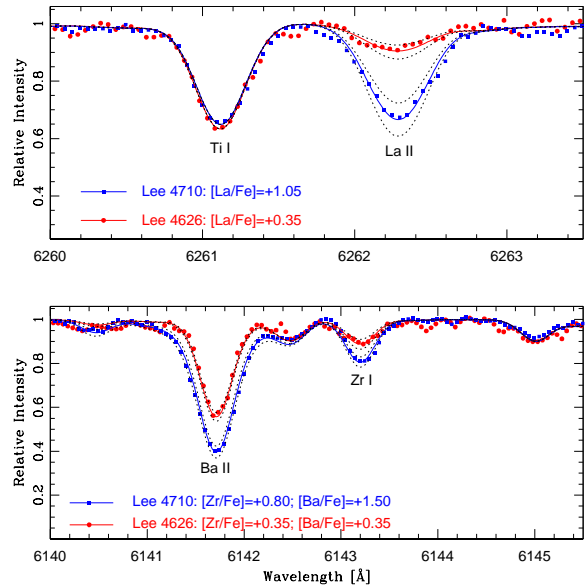}
  \end{center}
  \caption{Synthetic spectra of Ti, La, Ba, and Zr fit to the stellar spectra of Lee~4626 and Lee~4710.} 
  \label{fig:spec}
\end{figure}

The stars \object{Lee~4710} and \object{Lee~4626} were observed with FLAMES at VLT 
(ID: 088.D-0026(A)), using the gratings HR13, HR14, and HR15, which offer 
moderate to high spectral resolving power ($R = 22\,500, 17\,700,$ and $19\,300$, 
respectively) and a wavelength coverage ranging from 6100 to 6800\,\AA~ 
(including gaps). The data reduction was carried out with the girBLDRS pipeline
(http://girbldrs.sourceforge.net/), which performs bias subtraction, 
flat-fielding, wavelength calibration, and object extraction. Sky subtraction 
and telluric contamination removal were performed using the IRAF tasks 
{\it skysub} and {\it telluric}. Multiple exposures taken a few days apart were combined using the 
IRAF task {\it scombine}, achieving a signal-to-noise ratio (S/N) of $\sim75$ 
at 6200\,\AA.  Two sample spectral regions of Lee~4710 and the 
comparison star Lee~4626, which has similar atmospheric parameters, are 
contrasted in Fig.~\ref{fig:spec}.

Atmospheric parameters were adopted from C14, which employed 
40 FeI and 6 FeII lines.  A more detailed description of the 
parameter determinations can be found in C14.  The radial 
velocities of both stars are given in Table~\ref{Tab:abundances}.  The 
velocity values are in agreement with the systemic velocity and dispersion
found by \citet{Koch2008}, which suggests that both stars are true cluster
members.

The abundance analysis was performed using the local thermodynamic equilibrium 
(LTE) line analysis code MOOG \citep[][version 2014]{Sneden1973} and plane 
parallel, $\alpha$--enhanced 1D ATLAS9 model atmospheres.  The abundance of all 
elements except Fe were computed using spectrum synthesis. The line list 
was downloaded from VALD \citep{VALD2000} and updated with atomic data from 
NIST (http://www.nist.gov/pml/data/asd.cfm) and the Kurucz database 
(http://kurucz.harvard.edu).  A molecular CN line list from \citet{Sneden2014} 
was also included.  Since we could not measure C or N independently, we 
assumed [C/Fe]$=-0.5$ and [N/Fe]$=+0.7$ following 
\citet{Carretta2013}, and adjusted the [N/Fe] to match the strength 
of nearby CN features in our spectra.

The full wavelength range was inspected for lines useful for the chemical 
analysis, with an emphasis on identifying lines of elements predominantly 
produced by the s-process. Only clean lines without significant blends were
measured, and the line lists included hyperfine splitting for ScI and II, 
BaII, LaII, and EuII. Finally, we confirmed that our line list could reproduce the spectral features in Arcturus using the atomic data provided by 
\citet{Koch2008} and \citet{McWilliam2008}. The final abundances can be 
found in Table~\ref{Tab:abundances}. Despite not including NLTE corrections, we expect $\Delta$[X/Fe]$_{\rm NLTE}$ to be $\la$0.10 dex (e.g., see Figures 8-9 in T14). Note that for elements in which we 
derived abundances from both neutral and singly ionized transitions (e.g., Sc 
and Ti), the abundances given Table~\ref{Tab:abundances} represent average 
values.  As seen from Table~\ref{Tab:abundances}, the two stars have very 
similar stellar parameters and lighter element ($Z<30$) abundances.  However, 
the two stars have significantly different heavy element ($Z\ga30$) abundances.
This effect is illustrated in Fig.~\ref{fig:spec}. From C14 and current measuments, we know that Lee~4710 is a primordial and Lee~4624 an intermediate population star. Like most GC stars, 
Lee~4626 shows a [Ba/Eu]$<0$ and $\alpha-$element enhancement of 
$\sim0.4$\,dex, which is consistent with enrichment by Type II supernovae.

\subsection{Abundance Uncertainties}
The abundances and number of measured lines are listed in 
Table~\ref{Tab:abundances}.  To assess the fitting uncertainties, all 
abundances have been measured separately by the authors, and we generally 
obtained a good agreement ($\pm 0.05$\,dex) on a line-by-line basis. 
This uncertainty in the profile fitting is propagated together with 
uncertainties stemming from stellar parameters and the line list. 
The uncertainties of the abundances from individual lines were calculated by varying each star's parameters by the model atmosphere uncertainties reported in C14 and Table~\ref{Tab:abundances}. The final uncertainty in [X/Fe] stems from the stellar parameters added in quadrature together with the uncertainty from line fitting and atomic data ($\pm0.05$\,dex).

\begin{table}[!ht]
\caption{Stellar coordinates, parameters, and abundances.}
\label{Tab:abundances}
\begin{tabular}{l c c c c }
\hline
\hline
Coordinates  &  Lee~4626  &    &     Lee~4710   &  \\
\hline
Ra   & 00~23~10.75  &   & 00~23~10.75   &   \\
Dec &  -71~56~01.0  &   & -71~59~15.0  &   \\
RV[km/s] &  -13.1$\pm2.0$ & & -17.6$\pm1.5$&   \\
\hline
\multicolumn{5}{c}{Parameters }  \\
\hline
$T_{\textrm{eff}}^c$[K]  &  4500  & $\pm50$  & 4475  & $\pm50$  \\
log$g^c$  & 1.73  & $\pm0.20$  & 1.60  & $\pm0.20$  \\
$[$Fe/H$]^c$ & -0.81  & $\pm0.15$  & -0.78  &  $\pm0.15$ \\
$\xi^c$[km/s] & 1.70  & $\pm0.25$  & 1.65  &  $\pm0.25$ \\
\hline
\multicolumn{5}{c}{Elemental abundances}\\
$[$X/Fe$]^s$ & Lee~4626 & No. lines & Lee~4710 & No. lines \\
\hline
O    &    0.40$\pm0.14$   & 1  &     0.35$\pm0.17$      & 1 \\
Na   &    0.60$\pm0.14$	  & 2  &     0.13$\pm0.17$      & 2 \\
Mg   &    0.30$\pm0.14$	  & 2  &     0.30$\pm0.13$       & 2 \\
Al   &    0.59$\pm0.15$   & 2  &     0.38$\pm0.15$      & 2 \\
Si   &    0.3$\pm0.14$	  & 3  &     0.28$\pm0.14$      & 3 \\
Ca   &   0.39$\pm0.16$	  & 4  &     0.30$\pm0.19$       & $5^*$ \\
Sc   &   0.23$\pm0.13$	  &$5^*$&    0.17$\pm0.16$      & $4^*$ \\
Ti   &   0.42$\pm0.15$	  &$8^*$&    0.33$\pm0.18$      & $8^*$ \\
V    &   0.37$\pm0.15$	  & 6  &     0.35$\pm0.17$      & $8^*$ \\
Cr   &   0.20$\pm0.14$	  & 2  &     0.20$\pm0.15$       & 1 \\
Co   &   0.40$\pm0.14$	  & 2  &     0.30$\pm0.14$       & 2 \\
Ni   &   0.07$\pm0.14$	  & 6  &     0.02$\pm0.19$      & 7 \\
Zn   &   $<0.85$  	  & 1  &     $<0.95$   		& 1 \\
Y    &    --	 	  & -- &     0.75$\pm0.18$      & 1 \\
Zr   &   0.37$\pm0.14$	  & 3  &     0.83$\pm0.17$      & 4 \\
Ba   &   0.35$\pm0.24$	  & 1  &     1.03$\pm0.20$     & $2^*$ \\
La   &   0.35$\pm0.19$	  & 2  &     1.14$\pm0.14$      & $6^*$ \\
Ce   &   --     	  &  --&     1.00$\pm0.17$      & 1 \\ 
Pr   &  $<0.6$  	  & 1  &     $<0.9$   		 & 2 \\
Nd   &    --	 	  & -- &      $<1.1$  		 & 2 \\
Eu   &    0.50$\pm0.19$	  & 1  &     0.50$\pm0.17$       & 1 \\
\hline  
\hline
\end{tabular} 
\begin{itemize}
\item[$^c$]Values from C14.
\item[*]Uncertain due to line blends.  These measurements were given half weight in the final averaged abundance listed.
\item[s]Solar abundances from \citet{Asplund2009}.
\end{itemize} 
\end{table}

\section{Discussion and Conclusion}
\label{sec:concl}
Although most GC stars, including those in 47~Tuc, tend to show a heavy 
element composition that is dominated by the r-process, previous 
studies have found that 47~Tuc stars exhibit some s-process enrichment as 
well.  Furthermore, 47~Tuc's s-process abundances are similar to the values
found in halo stars with comparable metallicity \citep{James2004}.  However,
since the majority of 47~Tuc stars are not strongly s-process enhanced and
do not show significant star-to-star scatter for the neutron-capture 
elements (T14, C14), the high [X/Fe] ratios exhibited
by Lee~4710 for elements heavier than the Fe-peak make the star a clear outlier
in the cluster distribution.  The low occurrence of s-process enhanced stars 
in 47~Tuc favors an event such as binary mass transfer as the preferred 
enrichment scenario for Lee~4710.

According to \citet{Bisterzo2014}, more than 70\% of Y, Zr, Ba, La, and Ce are 
created by the s-process, and 94\% of Eu is created by the r-process at the time the solar system formed.  Therefore, finding
a low-mass star with enhanced [Ba/Eu], in systems with [Fe/H]$>\sim-2$, suggests
that a significant fraction of the star's heavy elements were produced by
the s-process.  Similarly, the large [X/Fe] ratios found for elements with 
Z$\ga$30 in Lee~4710 (see Table~\ref{Tab:abundances}) lead us to believe 
that this star experienced significant s-process pollution from an external 
source.  Since most GCs exhibit [Ba,La/Eu]$<0$ 
\citep[e.g., see][their Figure 6]{Gratton2004}, the heavy elements in these
clusters were likely dominated by r-process enrichment from supernovae.  Thus,
GC stars such as Lee~4710 are relatively rare to find among monometallic GCs.

\subsection{ A new s-process rich star in 47~Tucanae}\label{sec:sprocess}

47~Tuc is one of the most metal-rich clusters in which s-process dominated stars have been found.   Here, we report the detection of significant enrichment ($\langle$[s-process/Fe]$\rangle$=0.96) of several elements in the RGB star Lee~4710 that are likely produced by an s-process site in 47~Tuc.  Lee~4710 is the most s-process-rich star yet found in 47~Tuc.  In this section, we compare the abundance of neutron-capture elements in Lee~4710 to other red giants in the cluster \citep[e.g., W06;][]{Dorazi2010} and to the barium star \object{HD~5424}.

\begin{figure}
  \begin{center}
    \includegraphics[width=0.98\linewidth]{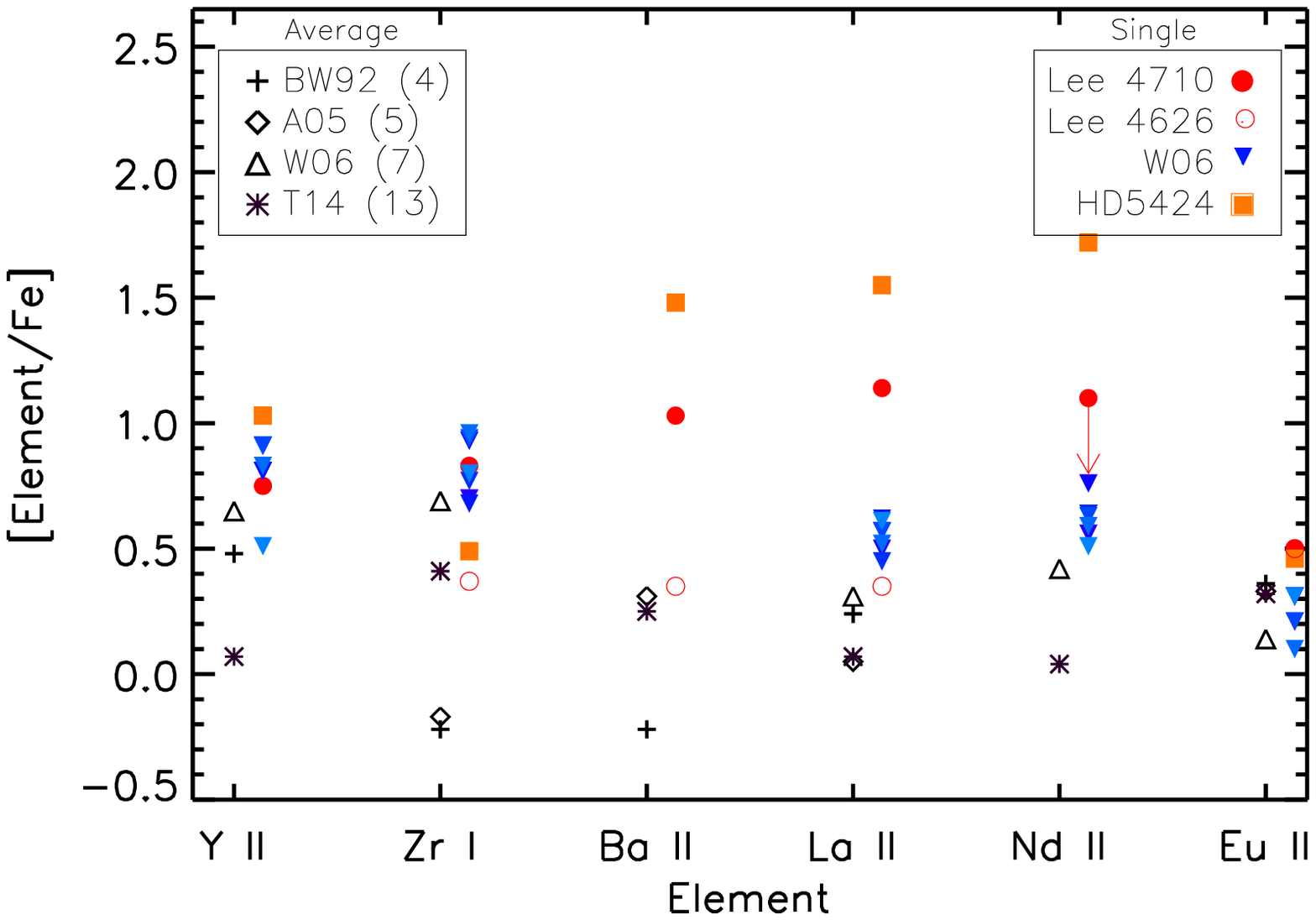}
  \end{center}
  \caption{Averaged abundances of 47~Tuc from \citet[][BW92]{BW1992}, \citet[][A05]{Alves2005}, \citet[][W06]{Wylie2006}, and \citet[][T14]{Thygesen2014} compared to single star abundances of: Lee~4710, Lee~4626, AGB stars from W06, and a Ba-star \citet{Allen2006}.  
All studies have been shifted to a common solar reference scale.} 
  \label{fig:average}
\end{figure}

Figure~\ref{fig:average} compares the scatter of heavy element abundances
derived from single stars to the scatter of the cluster average abundances 
determined in different studies.  Figure~\ref{fig:average} also includes a 
comparison with the similar metallicity ([Fe/H]$=-0.55$) Ba-star HD~5424 from
\citet{Allen2006}.  We find most analyses to be in agreement that the average 
[Eu/Fe] abundance (r-process) is similar to that 
of [Ba/Fe] and [La/Fe] (s-process).  These data suggest that, at least on 
average, stars in 47~Tuc are not strongly s-process enhanced.  
Figure~\ref{fig:average} indicates that some star-to-star scatter might 
exist for the lighter s-process elements (e.g., Y and Zr spanning 1.2\,dex).
While the cluster average abundance of [Zr/Fe] from \citet{BW1992}, \citet{Alves2005}, W06, and T14 varies by nearly a factor of ten, a close examination of the individual 
star abundances in their samples does not support an intrinsic dispersion in 
[Zr/Fe].  Within each study, the star-to-star [Zr/Fe] scatter is comparable
to that of the heavier neutron-capture elements for which the averages between
studies are in better agreement.

\begin{figure}
  \begin{center}
    \includegraphics[width=0.98\linewidth]{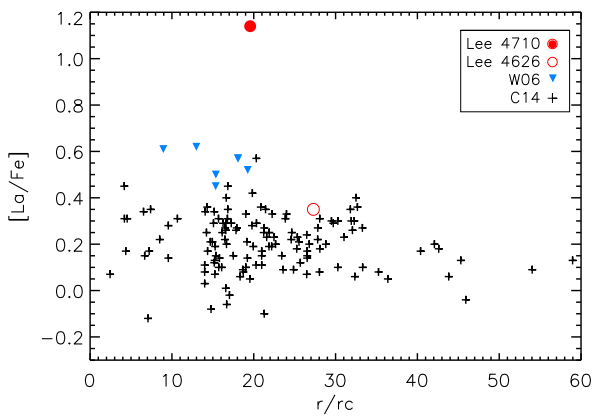}
  \end{center}
  \caption{Lanthanum abundances in 47~Tuc as a function of core radius. Comparison samples are from W06 and C14.}
  \label{fig:radialLa}
\end{figure}

Interestingly, Figure~\ref{fig:average} shows that Lee~4710 is the most 
s-process enhanced star yet found in the cluster. 
It seems unlikely that any intrinsic heavy element dispersion in 47~Tuc is due 
entirely to binary mass transfer.  Instead, the star-to-star [Ba,La/Fe] 
dispersion found for stars with generally low [Ba,La/Eu] ratios may be a 
reflection of primordial variations and/or incomplete mixing of gas within the 
early cluster environment.  In fact, heavy element dispersions ranging
from a factor of $\sim2-6$ due to primordial r-process inhomogeneities alone 
have already been detected in some clusters \citep{Roederer2011}.  However, it is 
interesting to note that our observations of Lee~4710 increase the number of 
47~Tuc stars with some s-process enhancement to nine stars, and all of them 
are located $>9$ core radii from the cluster center\footnote{Only 10/170 stars studied are located inside 9 core radii.  Therefore, the lack of s-rich stars $<9r/r_c$ may not be significant.} 
(see Fig.~\ref{fig:radialLa}). 

\subsection{S-process origin of Lee~4710}\label{sec:Lee4710}

Since Lee~4710 exhibits larger [Ba,La/Fe] excesses compared to [Y,Zr/Fe], we expect that the star was polluted by the main s-process operating in a low to intermediate mass AGB star.  We can 
rule out that Lee~4710 is a thermally pulsing AGB star given both its location
on the color-magnitude diagram (2.5 mag. below the RGB-tip) and the fact that
as a co-eval cluster member it would not have a high enough mass to be an
intrinsic Ba-star.  Thus, we believe that the s-process enrichment originates 
from a former AGB binary companion (see also 
Sect.~\ref{sec:binary}).  In order to better trace the possible origin of 
the s-process enrichment of Lee~4710, we can compare the star's stellar 
abundance pattern to theoretical yields from similar metallicity AGB stars
of different masses. In this connection we note that detailed, isotopic abundances of C and N in Lee~4710 would help us unveil the nature of its enrichment or mixing events, since dredge-up episodes would increase the $^{13}$C and N abundances and lead to a low C/N-ratio.

\begin{figure}
  \begin{center}
    \includegraphics[width=0.98\linewidth]{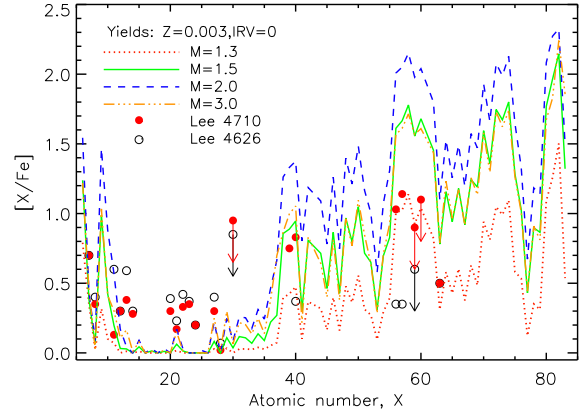}
  \end{center}
  \caption{Abundances of Lee~4710 and Lee 4624 compared to AGB yield predictions
  with fixed metallicity (Z=0.003), no rotation, and four different masses (in $M_{\odot}$ -- see legend).}
  \label{fig:yields}
\end{figure} 

Figure~\ref{fig:yields} shows a comparison of the heavy element abundances 
derived for Lee~4626 (s-process poor) and Lee~4710 (s-process rich) against 
theoretical yields from the F.R.U.I.T.Y database \citep{Cristallo2011} for 
M$=1.3, 1.5, 2.0$, and $3.0 M_{\odot}$ AGB stars.  Using these data, we find that 
for Lee~4710 the abundance pattern of elements with $Z>40$ is best fit by the 
1.3$M_{\odot}$ model.  The $1.5, 2.0$, and $3.0M_{\odot}$ AGB models 
produce [X/Fe] ratios that are $>$0.5 dex too large to match the observations.
Although we note that the present-day surface composition of Lee~4710 may
not entirely reflect the abundance pattern of the material accreted while the
star was on the main-sequence, any dilution of accreted material due to RGB
evolution is unlikely to change the heavy element abundances by 0.5\,dex or more, and we therefore believe that a low-mass AGB star ($<$1.5$M_{\odot}$) is most likely to provide the best fit (even after some pollution or dilution of the transferred material).
The [Y/Fe] and [Zr/Fe] abundances are also enhanced in Lee~4710 compared to 
Lee~4626, and these elements are better fit by the 1.5$M_{\odot}$ AGB model
compared to the more massive AGB models.  However, these elements may also have
a production component from the weak neutron-capture processes and therefore,
at least in 47~Tuc, the lighter elements may not be as useful for constraining the masses of 
individual AGB polluters.

\subsubsection{Light vs heavy s-process abundances}\label{sec:lightheavy}

A simple tracer of s-process production is the ratio of the heavy-to-light
elements, which is annotated as [hs/ls] where hs is the average of [Ba/Fe],
[La/Fe], [Nd/Fe], and [Sm/Fe] and ls is the average of [Sr/Fe], [Y/Fe], 
and [Zr/Fe].  As a rule of thumb, the lighter s-process elements, such as 
Sr -- Zr, can be created by the weak s-process 
\citep[e.g.,][]{Pignatari2010} in more massive stars.  The weak s-process
is predicted to produce moderate amounts of light s-nuclei and only
a small amount of the heavy s-elements.  This in turn results in a 
negative [hs/ls].  However, such a negative ratio can also be produced
by AGB stars if they have not undergone many pulses or if the 
seed-to-neutron ratio is high (i.e., a high metallicity).  Similarly,
an AGB star that experiences many thermal pulses and dredge-up episodes or 
has a low seed-to-neutron ratio (low metallicity) will produce a positive 
[hs/ls] ratio.

\citet{Wylie2006} find a [hs/ls] of $-0.4$ to $-0.09$, which on 
our abundance scale (and [Fe/H]) corresponds to $-0.35$ to $+0.05$. These 
values match those we derive for Lee~4626, which shows a normal La abundance.
On the other hand, Lee~4710 has a [hs/ls] of +0.33.  This large value 
almost matches the total [hs/ls]-ratio of a 1.3$M_{\odot}$ AGB star (Z=0.006) 
after all dredge-ups (a total of four episodes) have been completed. We note that if
the metallicity was slightly higher, then the observed [hs/ls] of Lee~4710
would agree with the 1.3$M_{\odot}$ AGB (Z=0.003) yield after just two 
dredge-up episodes. With the high [hs/ls] of Lee~4710, the 
weak s-process is therefore not likely to explain the formation of this star, 
whereas a few or even all dredge-up episodes of a 1.3$M_{\odot}$ AGB can
plausibly explain the heavy element enrichment found in Lee~4710.

\subsection{Binarity in GCs}\label{sec:binary}

Understanding the mechanism responsible for the s-process pattern of star Lee~4710 will shed light on its chemical enrichment and indirectly may provide 
some insight regarding dynamical differences between the inner and outer 
regions of 47~Tuc.  A commonly adopted explanation for the creation of 
extrinsic Ba-stars is mass transfer in a binary system 
\citep[e.g.,][W06]{McClure1980,Dorazi2010}.  In this scenario, a more massive 
companion produces s-process elements through third dredge-up episodes that
are later accreted onto the surface of the lower mass, and more slowly 
evolving, companion.  The old age of 47~Tuc 
\citep[11.75$\pm$0.25 Gyr;][]{Vandenberg2013} stars provides more than enough
time for a $1.3M_{\sun}$ star to evolve through the AGB phase and be 
responsible for the heavy element abundance pattern of Lee~4710.

As noticed by \citet{Dorazi2010}, four out of five Ba-rich stars in their 
sample ($\sim$1200 stars) are primordial population stars, similar to Lee~4710.
This finding is consistent with the greater predicted survival rates for 
binaries in the primordial population of 47~Tuc, and would allow the binary 
system to survive long enough for the original primary star to reach the AGB
phase and transfer s-process enhanced material onto its companion.  The fact 
that all nine s-process enriched stars found in 47~Tuc are far from the center 
of the cluster suggests that the outer regions may provide more favorable
conditions for binary systems to survive and for mass transfer to occur.
This scenario is supported by \citet{Hong2015} who claims that at an age of 5 Gyr, when the putative AGB companion transferred mass, 75\% of the binaries in the first generation population survived. At 47~Tuc's current age, $\sim 60$\% of these still survive.

Although Omega Cen is known to have a large population of stars highly 
enriched in s-process elements, [s-process/Fe]$>1.0$ 
\citep{Johnson2010,Marino2011}, large s-process enhancements are an uncommon 
characteristic among most GCs. Enhancements in s-process elements in GCs 
typically do not exceed [Ba,La/Fe]=0.6\,dex and have been found only in a 
limited number of more metal-poor clusters ([Fe/H]$<-1.2$): M4 
\citep{Shingles2014}, NGC~1851 \citep{Yong2008}, M22 \citep{Marino2009}, M2 
\citep{Yong2014}, and NGC~6864 \citep{Kacharov2013}.
From published samples of s-process elements ($\sim$200 stars) we find that 
nine stars are s-process enriched in 47~Tuc, i.e. $\sim$4\%$\pm$2\%. Within 
the errors, this number is consistent with 47~Tuc's small binary fraction 
(1.8$\%$$\pm$0.6$\%$, see \citealt{Milone2012}).  However, there are two caveats
that prevent assessing whether mass transfer in a binary system is the only 
mechanism responsible for the s-process rich stars in 47~Tuc. Firstly, \citeauthor{Milone2012} determined the fraction of binaries in 47~Tuc within 2.5 arcmin; 
therefore, in the outermost regions where the s-process rich stars have been 
found, the binary fraction could be different. \citet{Hong2015} suggest that the first generation binary fraction increases as a function of distance from the center. Secondly, the current binary 
fraction could be slightly smaller than it was when the s-rich stars were 
produced due to dynamical evolution of the cluster. 

Multi-epoch spectroscopic follow-up with several months between the observation is needed to determine whether some of these nine s-process rich stars currently belong to a binary system with an 
unseen WD or post-AGB companion.  These observations should also be 
accompanied by a determination of C, N, and $^{12}$C/$^{13}$C abundances that 
may help to further constrain any AGB mass transfer scenarios.

\begin{acknowledgements}
We thank the referee for useful comments.
M.~J.~Cordero and C.~J.~Hansen were supported by Sonderforschungsbereich SFB 881 "The
 Milky Way System" (subproject A4, A5, and A8) of the German Research Foundation (DFG) and C.J.H. also by a research grant (VKR023371) from VILLUM FONDEN.
C.I.J. acknowledges support from the Clay Fellowship, administered 
by the Smithsonian Astrophysical Observatory. C.J.H. thanks Thomas Masseron for a fruitful discussion.

\end{acknowledgements}

\bibliographystyle{apj}

\end{document}